\documentclass[11pt,a4paper]{article}
\pdfoutput=1
\usepackage{jheppub}
\usepackage[utf8]{inputenc}
\usepackage[english]{babel}
\makeatletter
\usepackage{amsmath,amssymb}
\usepackage{esint}
\usepackage{graphicx}
\usepackage{caption}
\usepackage{subcaption}
\usepackage{float}
\usepackage{mathtools}
\usepackage{xcolor}  
\usepackage{enumitem}

\usepackage{comment}

\DeclareMathOperator{\re}{Re}
\DeclareMathOperator{\im}{Im}
\newcommand{\Tr}{\text{Tr}}

\title{Entanglement of a chiral scalar on the torus}
\author[a]{Nicolás Abate,}
\author[b]{David Blanco,}
\author[b]{Alan Garbarz,}
\author[b]{Mateo Koifman}
\author[b]{and Guillem P\'erez-Nadal}
\affiliation[a]{Centro Atómico Bariloche, CONICET, and Instituto Balseiro\\8402 Bariloche, Río Negro, Argentina}
\affiliation[b]{Universidad de Buenos Aires, Departamento de F\'isica and IFIBA - CONICET\\
1428 Buenos Aires, Argentina}
\emailAdd{dblanco@df.uba.ar, guillem@df.uba.ar}

\abstract{
We compute the entanglement entropy of an interval for a chiral scalar 
on a circle at an arbitrary temperature. We use the resolvent method, which involves expressing the entropy in terms of the resolvent of a certain operator, and we compute that resolvent by solving a problem that entails finding an analytic function on the complex torus with certain jump conditions at the interval. The resolvent is relevant by itself, since it can be used to compute any function of the reduced density matrix. We illustrate that by also computing all the Rényi entropies for the model.}

\begin{document}

\maketitle

\section{Introduction}\label{sect:1}

In recent years, substantial insights into certain conceptual issues of Quantum Field Theory (QFT) have emerged through the study of quantities related to entanglement, such as the entanglement entropy.
Initially explored in connection with the challenge of identifying a quantum origin for the entropy of black holes \cite{PhysRevD.34.373,Sorkin:1984kjy}, this quantity has since been utilized to shed light on a variety of topics such as the renormalization group irreversibility \cite{Casini:2004bw,Casini:2012ei,Casini:2017vbe}, energy bounds \cite{Casini:2008cr,Blanco:2013lea,Faulkner:2016mzt,Balakrishnan:2017bjg} and symmetries \cite{Casini:2019kex,Casini:2020rgj,Casini:2021zgr}.

A common method used to compute entanglement entropies in QFT is the replica trick. This method consists of expressing the Rényi entropies of integer order $n\ne 1$ in terms of the partition function of the theory on a certain $n$-sheeted manifold. 
After computing this partition function, the entanglement entropy can be obtained by analytically extending the Rényi entropies to complex values of $n$ and letting $n\rightarrow 1$.

For free theories in Gaussian states, there is a more direct method. Indeed, in this case the entanglement entropy has the form $S=\Tr\, f(G_V)$, where $G_V$ is an operator constructed out of the two-point function. In order to compute functions of $G_V$, what we have to do is determine the spectral decomposition of this operator or, alternatively, its resolvent. Once we have any of these objects, we can compute the entanglement entropy, as well as other quantities associated with the reduced density matrix.

The two methods mentioned above, along with other techniques, have been used extensively to study entanglement in QFT in a variety of models (for instance, see \cite{Casini:2022rlv}). Among the free models, the case of a chiral fermion in two dimensions is remarkable since many quantities related to entanglement can be computed exactly for the vacuum and thermal states on the line and on the circle (a partial list of references is \cite{Casini:2009vk,Wong:2013gua,Blanco:2019xwi,Blanco:2019cet,Fries:2019ozf,Azeyanagi:2007bj,Herzog:2013py,Blanco:2021kzm,Abate:2022dyw,Abate:2023ldj}). For the scalar field results have been more elusive. In two dimensions the massless scalar field is ill-defined due to infrared divergences; one way to get rid of these divergences is to consider a chiral current, namely a null derivative of the scalar field. The entropy and modular Hamiltonian of a chiral current in the vacuum state reduced to two intervals were obtained in \cite{Arias:2018tmw}.

In this paper we use the resolvent method to compute the entanglement entropy of a chiral current for an interval on a circle at an arbitrary temperature.
This constitutes a new, previously unknown result. We compute the resolvent by first mapping the computation to a sort of Riemann-Hilbert problem (the problem of finding an analytic function on the complex torus satysfying certain jump conditions at the interval), similar to the problem introduced in \cite{Arias:2018tmw} for the eigenvectors of $G_V$. 
This newfound result for the resolvent can be useful to extensively study further aspects of entanglement in this model.

The paper is organized as follows. In section \ref{sect:2} we derive the expression for the entanglement entropy of a chiral scalar in terms of the resolvent of $G_V$. In section \ref{sec:2.5} we compute $G_V$ in the case of the torus, namely when the spatial manifold is a circle and the state is thermal. In section \ref{sect:3} we introduce our method to compute the resolvent, illustrating it in the simpler context of the plane, i.e., in the limit where the circle becomes a line and the temperature goes to zero. We then apply this method to the case of the torus
in section \ref{sect:4}.
The resolvent is used in section \ref{sect:5} to compute the entanglement and Rényi entropies, and
we conclude with a brief summary and discussion in section \ref{sect:6}.

\section{Entanglement entropy from the resolvent}\label{sect:2}

Consider a real quantum field $j$ on a manifold $\Sigma$ (which is to be regarded as space, the spacetime being ${\mathbb R}\times\Sigma$) satisfying commutation relations of the form
\begin{equation}
    [j(x),j(y)]=iC(x,y),\label{comm}
\end{equation}
where $C$ is antisymmetric, $C(y,x)=-C(x,y)$, and takes values in the real numbers. Suppose that the field is in a Gaussian state. The entanglement entropy of a region $V\subset \Sigma$ is \cite{Arias:2018tmw,Sorkin:2012sn}
\begin{equation}
    S = \Tr\left\{ \Theta\left( G_V-1/2\right)\left[ G_V \log G_V - (G_V-1)\log(G_V -1)\right]\right\}\,,\label{entropyC}
\end{equation}
where $G_V$ is an operator on the space of functions on $V$. This operator acts on a function $f$ via the equation $(G_Vf)(x)=\int_V dy\, G_V(x,y)f(y)$, with kernel
\begin{equation}
    G_V(x,y) = - i \int_V du\, \langle j(x)j(u)\rangle C^{-1}(u,y)\,, \label{GV}
\end{equation}
where $C^{-1}$ denotes the inverse of $C$, viewed also as an operator acting on functions on $V$.
The operator $G_V$ is generically not Hermitian, but 
it turns out to be diagonalizable with real spectrum. From the commutation relations it follows that the spectrum is contained in 
$\left(-\infty,0\right) \cup \left(1,+\infty\right)$ \cite{Arias:2018tmw}. Of course, these results are extrapolations from lattice computations, where the entanglement entropy (and more generally, the reduced density matrix) can be defined since there is a trace for the local algebras \cite{Witten:2018lha}.

\begin{figure}
    \centering
    \includegraphics[scale=0.7]{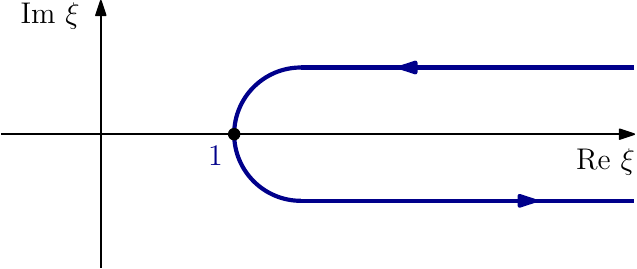}
    \caption{Contour of integration used for computing the entanglement entropy. The distance between the two horizontal stretches is infinitesimally small.}
    \label{fig:0}
\end{figure}

We introduce the resolvent of $G_V$,
\begin{equation}
    R(\xi) = \frac{1}{G_V-\xi}\,,\label{resdef}
\end{equation}
where $\xi$ is a complex number that is not in the spectrum of $G_V$. 
We can rewrite the entropy in equation (\ref{entropyC}) in terms of the resolvent. 
To do it, we use Cauchy's integral formula: if $g$ is an analytic function, then
\begin{equation}
    g(z)=-\frac{1}{2\pi i}\ointctrclockwise d\xi \, \frac{g(\xi)}{z-\xi}
\end{equation}
for any point $z$ enclosed by the contour. Moreover, for $z$ outside the contour the integral vanishes.
Defining
\begin{equation}
    g(\xi) = \xi \log \xi - (\xi-1)\log(\xi-1)\,, \label{g-entro}
\end{equation}
equation (\ref{entropyC}) takes the form $S = \Tr[\Theta(G_V-1/2) g(G_V)]$. Therefore, by Cauchy's integral formula, we can write
\begin{equation}
    S = -\frac{1}{2\pi i} \Tr \ointctrclockwise d\xi \, R(\xi) g(\xi)\label{entropyR}
\end{equation}
for any contour ${\mathcal C}$ with the following properties: (i) $g$ is analytic inside ${\mathcal C}$; (ii) ${\mathcal C}$ encloses the subset of values in the spectrum of $G_V$ that are greater than $1/2$, i.e., the interval $(1,+\infty)$; and (iii) ${\mathcal C}$ leaves out the rest of the spectrum of $G_V$, i.e., the interval $(-\infty,0)$.
An example of a contour satisfying these conditions is depicted in figure \ref{fig:0}.

Therefore, if we have the resolvent we can obtain the entanglement entropy (and more generally, any function of the reduced density matrix). Note that the discussion so far has been very general: we have considered a field satisfying generic commutation relations, and we have only assumed that it is in a Gaussian state. From now on we will be more specific.

\section{Chiral current on the torus}\label{sec:2.5}

We will take the field $j$ to be a chiral current on a circle of length $L$. This means that the commutator is given by equation (\ref{comm}) with
\begin{equation}
    C(x,y)=\pm\delta'(x-y)\,,\label{comC}
\end{equation}
and that the Hamiltonian is
\begin{equation}
    H = \frac{1}{2} \int_{-L/2}^{L/2} dx\, j^2(x)\,.
\end{equation}
We will refer to the upper and lower signs in equation (\ref{comC}) as the positive and negative chiralities respectively. This model can be obtained from the massless scalar field $\phi$ on the circle by defining 
\begin{equation}\label{jphi}
    j=\frac{1}{\sqrt 2}(\pi\pm\phi')\,,
\end{equation}
where $\pi$ denotes the canonical momentum; in other words, $j$ is proportional to the derivative of the time-evolved field with respect to 
the null coordinate $x^\pm=t\pm x$. On the other hand, the Gaussian state we will consider is a thermal state with inverse temperature $\beta$.

Let us compute the two-point function $\langle j(x)j(y)\rangle$ corresponding to these choices. We start with the Euclidean propagator $J$, defined by
\begin{equation}\label{euclidean}
    J(x,\tau)=\theta(\tau)\langle e^{H\tau}j(x)e^{-H\tau}j(0)\rangle+\theta(-\tau)\langle j(0)e^{H\tau}j(x)e^{-H\tau}\rangle
\end{equation}
for $\tau\in(-\beta,\beta)$. Clearly, by the translation invariance of the state we have
\begin{equation}
    \langle j(x)j(y)\rangle=\langle j(x-y)j(0)\rangle=J(x-y,\epsilon),
\end{equation}
where $\epsilon$ is a positive number to be sent to zero after smearing the field with a test function. The Euclidean propagator satisfies $J(x,\tau+\beta)=J(x,\tau)$ for $\tau\in(-\beta,0)$, so it can be extended to arbitrary values of $\tau$ as a periodic function. Due to this periodicity, we can think of $J$ as being defined on a torus of circumferences $L$ and $\beta$. Differentiating (\ref{euclidean}) with respect to $\tau$ and using the commutation relations, one obtains that $J$ satisfies the equation
\begin{equation}
    (\partial_x\mp i\partial_\tau)J=\delta(\tau)\delta'(x)\,.
\end{equation}
Identifying the torus with a complex torus via the map $(x,\tau)\mapsto x\mp i\tau$, this equation tells us that $J$ is analytic except for a
second-order pole at the origin with Laurent coefficient $-1/2\pi$ (the second part of this statement is shown by integrating $zJ(z)$ along a rectangular contour enclosing the origin and applying Green's theorem). These conditions completely determine $J$ except for an additive constant. Indeed, the difference between two functions satisfying these conditions is analytic and bounded throughout the complex plane (by periodicity), and hence constant by Liouville's theorem. A meromorphic function on the torus with a double pole at the origin and nowhere else is the Weierstrass elliptic function\footnote{In this paper we will make extensive use of the Weierstrass functions; see Appendix A of \cite{Blanco:2019cet} for a brief review and \cite{Pastras:2017wot} for a more detailed study of these beautiful functions.} $\wp$, which has Laurent coefficient $1$ for the double pole. The Euclidean propagator is thus
\begin{equation}
    J(z)=-\frac{1}{2\pi}\left[\wp(z)+c\right]\,,
\end{equation}
where $c$ is a constant. It follows that
\begin{equation}\label{2pt}
    \langle j(x)j(y)\rangle=-\frac{1}{2\pi}\left[\wp(x-y\mp i\epsilon)+c\right]\,.
\end{equation}
We can fix the value of $c$ by thinking of $j$ as arising from a massless scalar field as explained above. Equation (\ref{jphi}) implies
\begin{equation}
    \int_{-L/2}^{L/2}dx\,j(x)=\frac{1}{\sqrt 2}\int_{-L/2}^{L/2}dx\,\pi(x)=\sqrt\frac{L}{2}p,
\end{equation}
where $p=\pi_0$ is the Fourier component of $\pi$ with zero wave number or, in other words, the momentum of the zero mode of the scalar field. This mode is a free particle of unit mass, so for a thermal state we have $\langle p^2\rangle=1/\beta$ and in consequence
\begin{equation}
    \int_{-L/2}^{L/2}dx\,dy\,\langle j(x)j(y)\rangle=\frac{L}{2}\langle p^2\rangle=\frac{L}{2\beta}\,.
\end{equation}
On the other hand, from (\ref{2pt}) we obtain
\begin{equation}
    \int_{-L/2}^{L/2}dx\,dy\,\langle j(x)j(y)\rangle=\frac{L}{\pi}\zeta(L/2)-\frac{cL^2}{2\pi}\,,
\end{equation}
where $\zeta$ is the Weierstrass zeta function, defined by the conditions $\zeta'=\wp$ and $\zeta(z)-1/z\to 0$ as $z\to 0$. This function has the quasiperiodicity
$\zeta(z+P)=\zeta(z)+2\zeta(P/2)$, with $P$ any of the two periods of the complex torus ($L$ and $i\beta$). Comparing the above two equations and using the property $i\beta\zeta(L/2)-L\zeta(i\beta/2)=i\pi$, we finally determine the constant,
\begin{equation}
    c=\frac{1}{L}\zeta(L/2)+\frac{1}{i\beta}\zeta(i\beta/2)\,,
\end{equation}
which is real because $\zeta$ commutes with complex conjugation, $\zeta^*(z)=\zeta(z^*)$, and is odd, $\zeta(-z)=-\zeta(z)$.
This completes the calculation of the two-point function. The result is in agreement with the literature, see e.g.~\cite{EGUCHI1987308}.

Another aspect in which we will be less general than in the discussion of the previous section is in the type of subregion $V$. We will take this subregion to be a single interval, $V=(a,b)$. With this choice, we have{\footnote{Strictly speaking, the operator $C$ defined in (\ref{comC}) is not invertible because it annihilates the constant functions. Equation (\ref{C-1}) gives the unique antisymmetric kernel satisfying $CC^{-1}=1$. It also satisfies $C^{-1}C=1$ when acting on functions that vanish at the endpoints of the interval. For more details see \cite{Arias:2018tmw}.}}
\begin{equation}\label{C-1}
    C^{-1}(x,y)=\pm\frac{1}{2}\left[ \theta\left(x-y\right) - 
\theta\left(y-x\right)\right]\,.
\end{equation}
Substituting this equation and (\ref{2pt}) into (\ref{GV}) we find that the operator $G_V$ entering the formulas for the entanglement entropy acts on a function $f$ on $V$ as
\begin{equation}\label{GVG}
    G_Vf=(Gf)^\mp,
\end{equation}
where $Gf$ is a function on the torus, defined only on the complement of $\bar V$ (the closure of $V$), and the superscript $\mp$ means the limit as $V$ is approached from below/above, $F^\mp(x)=F(x\mp i\epsilon)$. The function is
\begin{equation}\label{G}
    (Gf)(z)=\int_{a}^b dy\,G(z,y)f(y)
\end{equation}
with
\begin{equation}\label{G(z,w)}
    G(z,w)=\pm\frac{1}{2\pi i}\left\{\zeta(z-w)-\frac{1}{2}\left[\zeta(z-a)+\zeta(z-b)\right]+c\left(w-\frac{a+b}{2}\right)\right\}\,.
\end{equation}
Note that $Gf$ is indeed defined on the torus, because $G(z,w)$ is periodic in $z$, but only on the complement of $\bar V$ because $G(z,w)$ has a first-order pole at $z=w$ and also at $z=a,b$.

For more general subregions of the circle, the expression for $C^{-1}$ becomes significantly more complicated than (\ref{C-1}) (see for instance the expression for two intervals, equation (5.2) in \cite{Arias:2018tmw}). This results in a considerably more complex form for $G_V$, which eventually leads to a problem for the resolvent that we have not yet been able to solve. This is why, in this paper, we focus only on the case of a single interval.

\section{Resolvent on the plane}\label{sect:3}

Our next step to compute the entropy is to find the resolvent $R$ of $G_V$, defined in (\ref{resdef}). If $f$ is a function on $V$, the function $Rf$ is the unique solution of the integral equation
\begin{equation}\label{reseq}
    G_VRf-\xi Rf=f.
\end{equation}
This equation looks complicated, but it can be mapped to a sort of Riemann-Hilbert problem, where one looks for an analytic function on the complex torus satisfying certain jump conditions at the interval. This Riemann-Hilbert problem can be solved more or less constructively. In order to introduce the method, in this section we concentrate on the case of the plane, $L,\beta\to\infty$. In the next section we will go to the more complex setting of the torus.

From now on we focus on the positive chirality. Note from equations (\ref{GVG})-(\ref{G(z,w)}) that the kernels $G_V(x,y)$ corresponding to the two chiralities are complex conjugates of each other. It then follows from (\ref{reseq}) that the resolvent of the negative chirality is related to that of the positive chirality by the equation $R_-(x,y;\xi)=R_{+}^*(x,y;\xi^*)$. In the limit $L,\beta\to\infty$, equation (\ref{G(z,w)}) for the positive chirality simplifies to
\begin{equation}
    G(z,w)=\frac{1}{2\pi i}\left[\frac{1}{z-w}-\frac{1}{2}\left(\frac{1}{z-a}+\frac{1}{z-b}\right)\right]\,,\label{GExplicit}
\end{equation}
because $\zeta(z)= 1/z$ when $L,\beta\to\infty$. The first step of the method is to study the properties of the corresponding function $Gf$.

\subsection{Properties of $Gf$}

\begin{figure}
    \centering
    \includegraphics[scale=0.55]{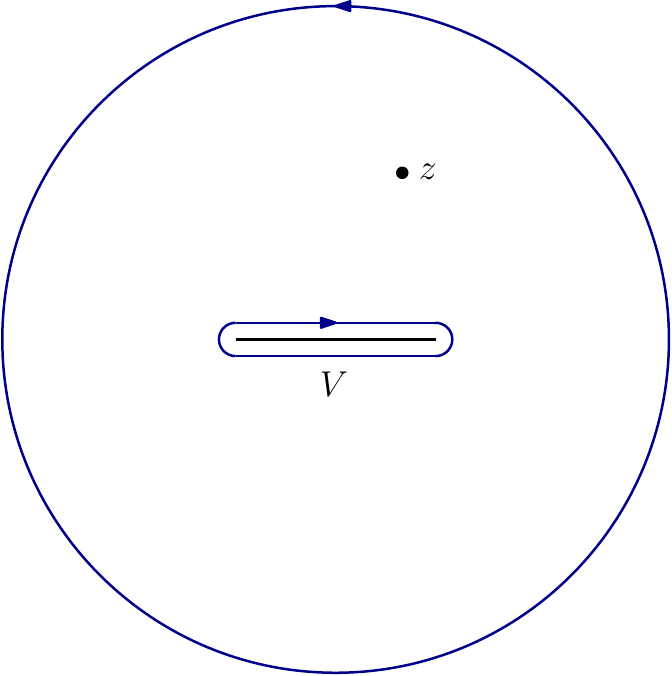}
    \caption{Integration contour for the first step in equation (\ref{uniqueness}).}
    \label{fig:contourF}
\end{figure}

From equations (\ref{G}) and (\ref{GExplicit}) it is clear that the function $Gf$ has the following properties.
\begin{enumerate}[label=(G\arabic*)]
    \item Analyticity: $Gf$ is analytic on $\mathbb{C}-\bar{V}$, where $\bar{V}$ is the closure of $V$.\label{g1}
    \item Asymptotic behavior: $(Gf)(z)={\mathcal O}(z^{-2})$ when $z\to\infty$\,.\label{infinityG}
    \item Behavior at the endpoints: $\lim_{z\to a}(z-a)(Gf)(z)=\lim_{z\to b}(z-b)(Gf)(z)<\infty$\,.\label{extremaG}
    \item Jump: $(Gf)^--(Gf)^+=f$\,.\label{saltoG}    
\end{enumerate}
The last property follows from the identity $1/(x-i\epsilon)-1/(x+i\epsilon)=2\pi i\delta(x)$. In fact, the converse statement is also true: if a function $F$ satisfies the conditions \ref{g1}-\ref{saltoG}, then $F=Gf$. Indeed, successive application of these conditions gives
\begin{alignat}{2}\label{uniqueness}
    F(z)&=-\ointctrclockwise dw\,G(z,w)F(w)=\ointctrclockwise_V dw\,G(z,w)F(w)\nonumber\\
    &=\int_{a}^b dy\,G(z,y)[F^-(y)-F^+(y)]=\int_{a}^b dy\,G(z,y)f(y)=(Gf)(z)\,.
\end{alignat}
In the first equality we use the analitycity of $F$ and integrate over the contour of figure \ref{fig:contourF}, which encloses the point $z$ and excludes the interval. In the second equality we use the asymptotic behavior of $F$, which ensures that the big circle in figure \ref{fig:contourF} does not contribute. In the third step we use the behavior near the endpoints, which implies that the contributions from the semicircles around the interval cancel out. And in the fourth step we use the jump condition. Therefore, the properties \ref{g1}-\ref{saltoG} completely characterize $Gf$.

\subsection{An equivalent problem}
\label{sec:equivalent}

Consider the function $S=GRf$. From the resolvent equation (\ref{reseq}) it follows that
\begin{equation}\label{Sres}
    S^-=\xi Rf+f\,.
\end{equation}
Taking this into account, the properties of $G$ imply the following properties of $S$.
\begin{enumerate}[label=(S\arabic*)]
    \item Analyticity: $S$ is analytic on $\mathbb{C}-\bar{V}$\,.\label{s1}
    \item Asymptotic behavior: $S(z)={\mathcal O}(z^{-2})$ when $z\to\infty$\,.\label{s2}
    \item Behavior at the endpoints: $\lim_{z\to a}(z-a)S(z)=\lim_{z\to b}(z-b)S(z)<\infty$\,.\label{s3}
    \item Jump: $S^--S^+=(S^--f)/\xi$\,.\label{s4}    
\end{enumerate}
In fact, there are no other functions with these properties. Indeed, the difference $\Delta$ between any such function and $S$ satisfies the conditions \ref{s1}-\ref{s4} with $f=0$, and hence, by the discussion at the end of the previous subsection,
\begin{equation}\label{Delta}
    \Delta=\frac{1}{\xi}G\Delta^-\,.
\end{equation}
Evaluating this equation just below the interval we find $(G_V-\xi)\Delta^-=0$, and hence $\Delta^-=0$ because $\xi$ is not in the spectrum of $G_V$. Substituting this result back in (\ref{Delta}) we finally obtain $\Delta=0$.

Therefore, the conditions \ref{s1}-\ref{s4} define a problem with a unique solution. Our strategy to compute the resolvent is to solve this problem and then substitute the solution into equation (\ref{Sres}).

\subsection{The solution}

Let us solve the problem \ref{s1}-\ref{s4}. 
What makes the solution not obvious is the fact that $S$ appears on the right-hand side of the jump condition \ref{s4}, so that the explicit value of the jump is not known. This suggests that we start looking for a solution $M$ of the difficult part of that condition, namely the jump condition with $f=0$,
\begin{equation}\label{jumpM}
    \frac{M^+}{M^-} = 1- \frac{1}{\xi}\,.
\end{equation}
The function $A=S/M$ then satisfies the jump condition
\begin{equation}\label{jumpA}
    A^--A^+=\frac{1}{1-\xi}\frac{f}{M^-}\,,
\end{equation}
which has the desirable property of not involving $A$ on the right-hand side. 
It is easy to find solutions of (\ref{jumpM}). For example, the function
 \begin{equation}
    \Omega(z) = \int_a^b \frac{dy}{z-y}=\log\frac{z-a}{z-b} \label{OmegaPlano}
\end{equation}
satisfies $\Omega^--\Omega^+=2\pi i$, so we can choose
\begin{equation}\label{M}
    M = e^{-ik\Omega}
\end{equation}
with
\begin{equation}
    k =\frac{1}{2\pi} \log \frac{\xi}{1-\xi}+\frac{i}{2}\,.\label{k}
\end{equation}
Note that $k$ is an analytic function of $\xi$ throughout the range of values of this variable (the complement of $(-\infty,0]\cup[1,\infty)$ in the complex plane), and takes values in the strip $\im k\in(0,1)$.
The function (\ref{M}) is 
analytic on $\mathbb{C}-\bar{V}$ and tends to $1$ at infinity. Moreover, $\exp[\pm i(\re k)\Omega]$ is bounded, because $\im\Omega\in(-\pi,\pi)$. From these properties, together with the properties of $S$, it follows that, besides the jump condition (\ref{jumpA}), the function $A=S/M$ satisfies \ref{s1}, \ref{s2} and the endpoint condition
\begin{equation}\label{endpointsA}
    \lim_{z\to a}\,(z-a)^\alpha A(z)=\lim_{z\to b}\,(z-b)^\beta A(z)=0
\end{equation}
for $\alpha>1+\im k$ and $\beta>1-\im k$. A first attempt at a solution to this problem is
\begin{equation}
    A_0=\frac{1}{1-\xi}G(e^{ik\Omega^-}f)\,,
\end{equation}
which satisfies all the conditions except for the endpoint condition at $b$.
In order to correct this ansatz, we subtract from $A_0$ the term responsible for its bad behavior near $b$, which comes from the last term in (\ref{GExplicit}). Doing that spoils the behavior at infinity, but this can be corrected by adding another term, which modifies the behavior near $a$ but without spoiling the corresponding endpoint condition. This gives
\begin{equation}
    A=\frac{1}{1-\xi}\tilde G(e^{ik\Omega^-}f)\,,
\end{equation}
where
\begin{equation}
    \tilde G(z,w)=G(z,w)+\frac{1}{4\pi i}\left(\frac{1}{z-b}-\frac{1}{z-a}\right)=\frac{1}{2\pi i}\left(\frac{1}{z-w}-\frac{1}{z-a}\right)\,.
\end{equation}
Therefore, the solution to the problem \ref{s1}-\ref{s4} is
\begin{equation}
    S=\frac{e^{-ik\Omega}}{1-\xi}\tilde G(fe^{ik\Omega^-}).
\end{equation}
One can check that this function indeed satisfies all the required properties. 

Substituting this solution into (\ref{Sres}) and using the identity $1/(x-i\epsilon)=1/x+i\pi\delta(x)$, where a principal value is implicit in the first term, we finally obtain the resolvent,
\begin{equation}
    R(x,y) = \frac{1}{\xi(1-\xi)} \left\{\frac{e^{-i k[ \omega(x) - \omega(y)]}}{2\pi i}\left(\frac{1}{x-y}-\frac{1}{x-a}\right)+\left(\xi-1/2\right)\delta(x-y)\right\}\,,\label{resolventeplano}
\end{equation}
where 
\begin{equation}
    \omega(x) = \log \left| \frac{x-a}{x-b}\right|
\end{equation}
and we have also used that $\Omega^-=\omega+i\pi$. 
This resolvent seems to treat very differently the endpoints $a$ and $b$, but this is just an appearance: if the above equation is rewritten in terms of $q=k-i/2$, the expression looks much more symmetric.

\subsection{Entanglement entropy on the plane}\label{sect:entroplano}

As a consistency check of the resolvent just obtained, let us use it to compute the entanglement entropy via equation (\ref{entropyR}). Since that equation involves a trace, we will ultimately set $x=y$, so it is convenient to expand the first term in (\ref{resolventeplano}) in powers of $x-y$,
\begin{alignat}{2}\label{expplano}
    R(x,y) =& \frac{1}{\xi(1-\xi)} \left\{\frac{1}{2\pi i}\left[\frac{1}{x-y}-ik\omega'(x)-\frac{1}{x-a}\right]+\left(\xi-1/2\right)\delta(x-y)\right\}\nonumber\\
    &+{\mathcal O}(x-y)\,.
\end{alignat}
The integral in (\ref{entropyR}) is performed along the contour depicted in figure \ref{fig:0}, which encloses the interval $(1,\infty)$. All terms in (\ref{expplano}) except for the second are analytic in $\xi$, so they do not contribute to the integral. We thus obtain
\begin{alignat}{2}\label{Splane1}
    S&=-\frac{1}{2\pi i}\int_{a+\epsilon}^{b-\epsilon}dx\,\lim_{y\to x}\ointctrclockwise d\xi\, R(x,y;\xi)g(\xi)\nonumber\\
    &=\frac{1}{(2\pi)^2i}\int_{a+\epsilon}^{b-\epsilon}dx\,\omega'(x)\ointctrclockwise d\xi\,\frac{k(\xi)}{\xi(1-\xi)}g(\xi)\nonumber\\
    &=\frac{1}{2\pi^2 i}\log\frac{\ell}{\epsilon}\ointctrclockwise d\xi\,\frac{k(\xi)}{\xi(1-\xi)}g(\xi)\,,
\end{alignat}
where $\epsilon$ is a cutoff used to regularize the integral and $\ell=b-a$ is the length of the interval $V$. The function $k(\xi)$ has a cut along the interval $(1,\infty)$, accross which it jumps according to $k^--k^+=-i$ , so we have
\begin{equation}
    \ointctrclockwise d\xi\,\frac{k(\xi)}{\xi(1-\xi)}g(\xi)=-i\int_{1}^\infty d\xi\,\frac{g(\xi)}{\xi(1-\xi)}=\frac{i\pi^2}{3}\,.
\end{equation}
Substituting into (\ref{Splane1}) we finally obtain
\begin{alignat}{2}
    S = \frac{1}{6}\log \frac{\ell}{\epsilon}\,, \label{entroPlano}
\end{alignat}
which is the universal result \cite{Holzhey:1994we} for the entanglement entropy of an interval when the theory is a conformal model on the line with central charge equal to $1/2$ (as is the case of the chiral current) and the global state is the vacuum.

\section{Resolvent on the torus}\label{sect:4}

Let us now compute the resolvent for arbitrary values of $L$ and $\beta$. As already mentioned, we focus on the positive chirality, which involves no loss of information because of the simple relation explained above with the negative chirality. We apply the method described in the previous section. The function $Gf$ behaves the same way near the endpoints of $V$ as its counterpart on the plane, and also satisfies the same jump condition. The analyticity property changes slightly because the complex plane ${\mathbb C}$ has to be replaced by the complex torus ${\mathbb T}$, and there is no condition at infinity because there is no infinity (one may say that the asymptotic property is replaced by periodicity, which is implicit in the fact that $Gf$ is a function on the torus). Therefore, we have the following properties.
\begin{enumerate}[label=(G\arabic*)]
    \item Analyticity: $Gf$ is analytic on $\mathbb{T}-\bar{V}$.\label{gt1}
    \item Behavior at the endpoints: $\lim_{z\to a}(z-a)(Gf)(z)=\lim_{z\to b}(z-b)(Gf)(z)<\infty$\,.\label{gt2}
    \item Jump: $(Gf)^--(Gf)^+=f$\,.\label{gt3}    
\end{enumerate}
Contrarily to the case of the plane, these properties do not characterize $Gf$ completely: if a function $F$ satisfies these three conditions, then so does $F+c$ for any constant $c$. In order to characterize $Gf$ completely, we need another property. The Weierstrass zeta function has a first-order pole at the origin with unit residue, and no other poles on the torus (the pole structure of $\zeta$ is periodic even though $\zeta$ itself is not).
Therefore, $G(z,w)$ has a pole at $z=w$ with residue $\pm 1/2\pi i$ (the sign depending on whether we view it as a function of $z$ or $w$), and two other poles at $z=a,b$, both with residue $-1/4\pi i$. On the other hand, it is a simple matter to check that $G(z,a)+G(z,b)=0$. All this implies
\begin{equation}
    \ointctrclockwise du\,G(z,u)G(u,w)=-G(z,w)+G(z,w)-\frac{1}{2}[G(z,a)+G(z,b)]=0
\end{equation}
for any contour enclosing the points $z$, $w$, $a$ and $b$. In consequence,
\begin{equation}
    \ointctrclockwise dw\,G(z,w)(Gf)(w)=0
\end{equation}
for any contour enclosing the point $z$ and the interval. This is the extra property we were looking for. Together with \ref{gt1}-\ref{gt3}, it completely characterizes $Gf$, i.e., if a function $F$ satisfies these four conditions then $F=Gf$. This is easily seen by an argument completely analogous to the one used in the case of the plane, equation (\ref{uniqueness}).

As a consequence of these properties and equation (\ref{Sres}), the function $S=GRf$ satisfies the following conditions.
\begin{enumerate}[label=(S\arabic*)]
    \item Analyticity: $S$ is analytic on $\mathbb{T}-\bar{V}$.\label{st1}
    \item Behavior at the endpoints: $\lim_{z\to a}(z-a)S(z)=\lim_{z\to b}(z-b)S(z)<\infty$\,.\label{st2}
    \item Jump: $S^--S^+=(S^--f)/\xi$\,.\label{st3}    
\end{enumerate}
Moreover, we have
\begin{equation}\label{st4}
    \ointctrclockwise dw\,G(z,w)S(w)=0
\end{equation}
for any contour enclosing the point $z$ and the interval. Again, the problem defined by these four conditions  has a unique solution by the same argument we gave around equation (\ref{Delta}).

Let us solve this problem. The strategy will be to find the most general solution to \ref{st1}-\ref{st3} and then impose condition (\ref{st4}). 
A solution to \ref{st1}-\ref{st3} is
\begin{equation}\label{S0}
    S_0=\frac{e^{-ik\Omega}}{1-\xi}H(e^{ik\Omega^-}f)\,,
\end{equation}
where $k$ is given by (\ref{k}) and
\begin{equation}
    \Omega(z)=\int_{a}^b dy\,\zeta(z-y)\,. \label{omegaToro}
\end{equation}
On the other hand, $(Hg)(z)=\int_{a}^b dy\,H(z,y)g(y)$ with
\begin{equation}\label{H(z,w)}
    H(z,w)=\frac{1}{2\pi i}\frac{\sigma(z-w+ik\ell)}{\sigma(z-w)\sigma(ik\ell)}\,,
\end{equation}
where $\ell$ is the length of the interval and $\sigma$ is the Weierstrass sigma function, defined by the conditions $\sigma'/\sigma=\zeta$ and $\sigma'(0)=1$. This function has the quasiperiodicity $\sigma(z+P)=-\exp[\zeta(P/2)(2z+P)]\sigma(z)$, it is analytic everywhere, and it has a zero at the origin and no other zeros on the torus (similarly to $\zeta$, the root structure of $\sigma$ is periodic even though $\sigma$ itself is not). Using these properties, one can easily check that $S_0$ indeed satisfies \ref{st1}-\ref{st3}.

The difference  $\Delta$ between two solutions of \ref{st1}-\ref{st3} satisfies the same conditions with $f=0$, which we will refer to as the homogeneous problem. A solution to this problem is
\begin{equation}\label{Delta0}
    \Delta_0(z)=e^{-ik\Omega(z)}\frac{\sigma(z-a+ik\ell)}{\sigma(z-a)}\,,
\end{equation}
as can be easily checked. In fact, the most general solution of the homogeneous problem is proportional to $\Delta_0$ with a constant coefficient. To see this, note first that the homogeneous problem is linear, so its solutions form a vector space. On the other hand, if $F$ is an analytic function on ${\mathbb T}-\bar V$ we have
\begin{equation}
    \frac{d}{dz}\ointctrclockwise dw\, G(z,w)F(w)=0
\end{equation}
for any contour enclosing $z$ and $V$. This is because $\partial_z G$ is periodic in $w$, so the left-hand side above is the integral of a periodic function, i.e., a function on the torus.  
This function is analytic on one of the sides of the contour (the side that does not contain $z$ and $V$). Since we are on the torus, both sides can be viewed as the interior of the contour,
so the integral must vanish. Therefore, the equation
\begin{equation}\label{map}
    \Delta\mapsto\ointctrclockwise dw\, G(z,w)\Delta(w)\,,
\end{equation}
with the contour enclosing $z$ and $V$, defines a linear map from the space of solutions of the homogeneous problem to the complex numbers. If the image of $\Delta$ by this map vanishes, then $\Delta$ satisfies \ref{st1}-\ref{st3} and (\ref{st4}) with $f=0$, and hence, by the uniqueness of the solution to that problem, $\Delta=GR0=0$. In other words, the map (\ref{map}) is injective, and hence the space of solutions of the homogeneous problem must have dimension 1.

We thus conclude that the most general solution to \ref{st1}-\ref{st3} is
\begin{equation}\label{solution}
    S=S_0+\lambda\Delta_0,
\end{equation}
with $\lambda$ a constant. It only remains to fix $\lambda$ by imposing (\ref{st4}). This gives
\begin{equation}
    \lambda=-\frac{\ointctrclockwise dw\, G(z,w)S_0(w)}{\ointctrclockwise dw\, G(z,w)\Delta_0(w)}\,,
\end{equation}
which after some algebra translates into
\begin{equation}\label{lambda}
    \lambda=-\frac{1}{2\pi i(1-\xi)\sigma(ik\ell)}\int_{a}^b dy\,\mu(y,k)e^{ik\Omega^-(y)}f(y),
\end{equation}
where 
\begin{equation}\label{mu}
    \mu(y,k)=\frac{I(y,k)}{I(a,k)}
\end{equation}
and
\begin{equation}\label{I}
    I(y,k)=\ointctrclockwise dw\,G(z,w)e^{-ik\Omega(w)}\frac{\sigma(w-y+ik\ell)}{\sigma(w-y)},
\end{equation}
with the contour enclosing $z$ and $V$. Note that $I$ is independent of $z$ by the discussion above. Note also that $I(a,k)$ is the image of $\Delta_0$ by the map (\ref{map}), so it is non-zero and hence $\mu$ is finite.
Equation (\ref{solution}), with $S_0$, $\Delta_0$ and $\lambda$ given respectively by (\ref{S0}), (\ref{Delta0}) and (\ref{lambda}), is the solution to the problem defined by conditions \ref{st1}-\ref{st3} and (\ref{st4}).


Substituting this result into (\ref{Sres}) we finally obtain the resolvent,
\begin{alignat}{2}
    R(x,y) = \frac{1}{\xi (1-\xi)} \Bigg\{&\frac{e^{-ik\left[ \omega(x)-\omega(y)\right]}}{2\pi i} \left[ \frac{\sigma(x-y+ik\ell)}{\sigma(x-y)\sigma(ik\ell)}-\mu(y,k) \frac{\sigma(x-a+ik\ell)}{\sigma(x-a)\sigma(ik\ell)} \right]  \nonumber\\
    &+\left(\xi-1/2\right) \delta(x-y)\Bigg\}\,,
    \label{resolventTorus}
\end{alignat}
where a principal value is implicit in the first term and
\begin{equation}
    \omega(x)=\log \left| \frac{\sigma(x-a)}{\sigma(x-b)}\right|\,.
\end{equation}
In obtaining these expressions we have used again the identity $1/(x-i\epsilon)=1/x+i\pi\delta(x)$, which in particular implies $\Omega^-=\omega+i\pi$. Equation (\ref{resolventTorus}) is one of the main results of this paper. Using this resolvent we can in principle compute any function of the density matrix, such as the entanglement entropy. We perform that computation in the next section.

Note that the above resolvent is very similar in structure to that of the plane, equation (\ref{resolventeplano}). In fact, it is easy to show that it reduces to the plane result in the limit $L,\beta\to\infty$, taking into account that $\sigma(z)= z$ in this limit. Indeed, if we are on the plane we can push the contour of integration in (\ref{I}) to infinity, so that the $y$-dependent factor in the integrand tends to $1$, thus making $I$ independent of $y$ and hence $\mu=1$. Substituting into (\ref{resolventTorus}) one straightforwardly recovers (\ref{resolventeplano}).

\section{Entanglement of a chiral current on the torus}\label{sect:5}

\subsection{Entanglement entropy}

\begin{figure}
    \centering
    \includegraphics[scale=0.50]{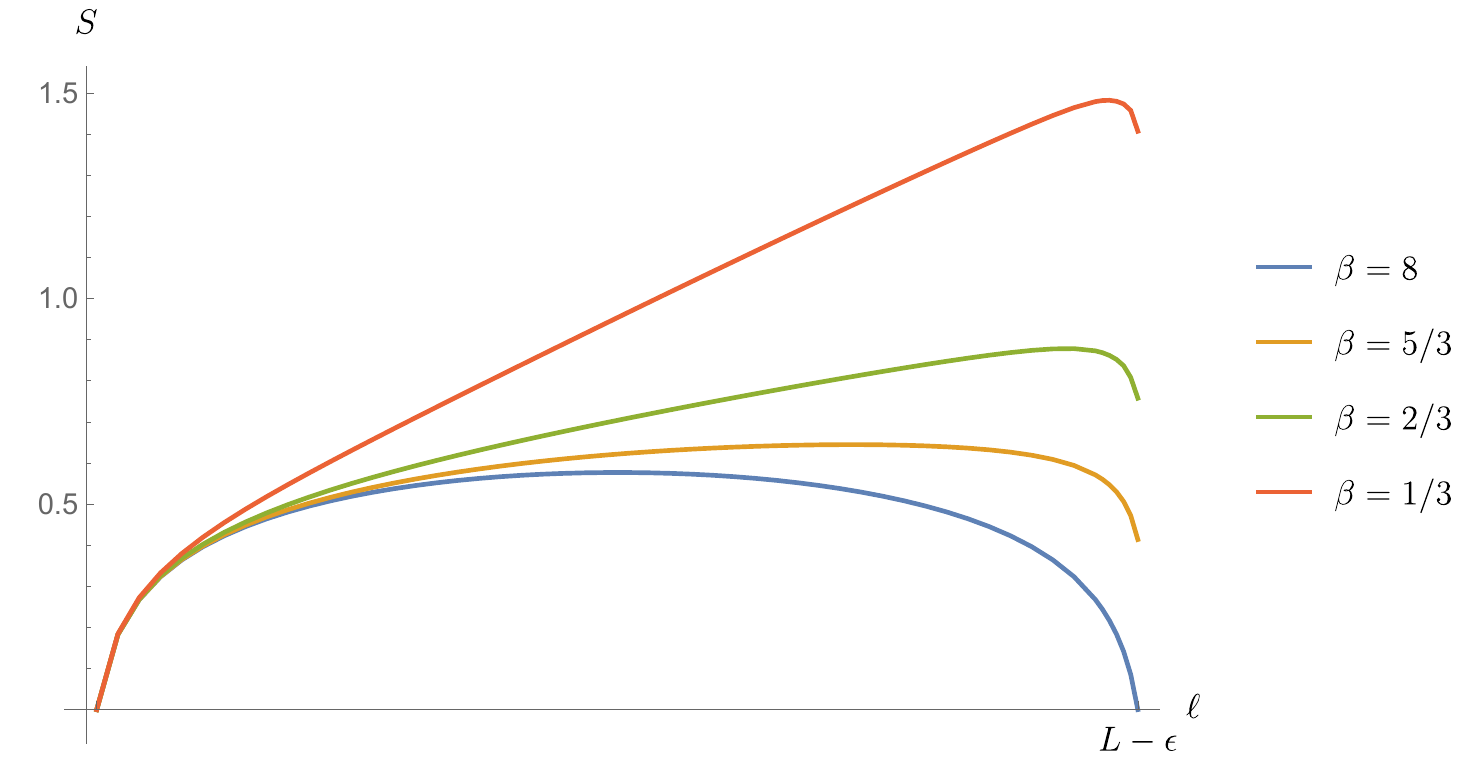}
    \caption{Entanglement entropy of an interval of size $\ell$ on the torus for $L=1$, $\epsilon = 1/100$ and various values of $\beta$. At high temperatures the entropy is almost linear because thermal effects dominate over entanglement, while for the vacuum state the entropy is symmetric with respect to the point $\ell = L/2$ because the field is in a pure state.}
    \label{fig:3}
\end{figure}

In this section we compute the entanglement entropy of the chiral current on the torus. The computation is similar to the one we did for the plane in section \ref{sect:entroplano}. We use equation (\ref{entropyR}) and integrate over the contour depicted in Figure \ref{fig:0}. Since we are eventually taking $x \rightarrow y$ we expand the resolvent (\ref{resolventTorus}) in powers of $x-y$,
\begin{alignat}{2}\label{exptoro}
    R(x,y)=\frac{1}{\xi(1-\xi)}\Bigg\{&\frac{1}{2\pi i}\left[\frac{1}{x-y}-ik\omega'(x)+\zeta(ik\ell)-\mu(x,k) \frac{\sigma(x-a+ik\ell)}{\sigma(x-a)\sigma(ik\ell)}\right]\nonumber\\
    &+\left(\xi-1/2\right) \delta(x-y)\Bigg\}+{\mathcal O}(x-y)\,.
\end{alignat}
Note the similarity with the case of the plane, equation (\ref{expplano}). Unlike that case, however, here we have three terms which are not analytic in $\xi$ inside the contour and hence contribute to the entropy: the second term, analogous to the only relevant term in (\ref{expplano}), and also the third and fourth. Thus, the entropy has a contribution analogous to the result for the plane, and an extra contribution coming from the third and fourth terms in (\ref{exptoro}),
\begin{equation}\label{EEtorus}
    S = \frac{1}{6} \log \frac{\sigma(\ell)}{\epsilon}+ \Delta S\,,
\end{equation}
with
\begin{equation}\label{DeltaS}
    \Delta S=\frac{1}{4\pi^2} \int_a^b dx\,\ointctrclockwise d\xi \frac{g(\xi)}{\xi (1-\xi)} \left[\zeta(ik\ell)-\mu(x,k) \frac{\sigma(x-a+ik\ell)}{\sigma(x-a)\sigma(ik\ell)} \right]\,.
\end{equation}
Defining
\begin{equation}
    \nu(x,k) = \mu(x,k) \frac{\sigma(x-a+ik\ell)}{\sigma(x-a)\sigma(ik\ell)}- \zeta(ik \ell)\label{nu}
\end{equation}
and changing the integration variable from $\xi$ to $k$, equation (\ref{DeltaS}) can be rewritten as
\begin{equation}\label{entropiatoroF}
    \Delta S=\frac{1}{2\pi} \int_{a}^{b} dx \, \int_{0}^{\infty} dk \, \left[ \nu(x,k) - \nu(x,k+i) \right] \left[ \frac{\log\left( e^{2\pi k} - 1 \right)}{e^{2\pi k}-1} - \frac{\log\left( 1 - e^{-2\pi k} \right)}{1 - e^{-2\pi k}} \right].
\end{equation}
Equations (\ref{EEtorus}) and (\ref{entropiatoroF}) give our final expression for the entanglement entropy of an interval for a chiral scalar on the torus. We have checked that $\Delta S$ is real and finite, so the divergent contribution to the entropy comes only from the first term in (\ref{EEtorus}). The integration in (\ref{entropiatoroF}) can be done numerically and so these results are also checked numerically. In figure \ref{fig:3} we present some plots of the entropy as a function of the size of the interval $\ell$ for $L=1$, $\epsilon = 1/100$ and several temperatures. We can see that at high temperatures the entropy is almost linear because thermal effects dominate over entanglement. On the other hand, at zero temperature, the entropy is symmetric with respect to the point $\ell = L/2$ because the field is in a pure state.

Lastly, as an explicit check, let us analyze the limits in which one of the periods of the torus goes to infinity, so that the torus becomes a cylinder; in these limits, the entropy is already known for any CFT. 
If $P$ is the period that remains finite, the relevant Weierstrass functions behave as (see Appendix A of \cite{Blanco:2019cet})
\begin{equation}
    \sigma (z) = \frac{P}{\pi} \sin \left(\frac{\pi z}{P}\right)e^{\frac{1}{6}\left(\frac{\pi z}{P}\right)^2}\qquad \zeta (z) = \frac{\pi}{P} \left[ \cot \left( \frac{\pi z}{P}\right) + \frac{1}{3} \frac{\pi z}{P} \right]\,. \label{weiersCylinder}
\end{equation}
Using these equations in (\ref{mu}) one can easily obtain{\footnote{As in the case of the plane, this is shown by pushing the integration contour in (\ref{I}) to infinity.}} an explicit expression for $\mu$,
\begin{equation}
\label{eq:mu}
\mu(x,k)=e^{-\frac{\pi^2}{3P^2}ik\ell(x-a)},
\end{equation}
which after substitution in (\ref{nu}) yields
\begin{equation}
    \nu(x,k) = \frac{\pi}{P} \left\{ \cot \left[ \frac{\pi \left( x-a\right)}{P}\right] - \frac{i\pi k \ell}{3 P}\right\}\,.
\end{equation}
Therefore, the difference $\nu(x,k) - \nu(x,k+i)$ appearing in (\ref{entropiatoroF}) is constant in $x$ and the integral can be computed,
\begin{equation}
    \Delta S=-
     \frac{\pi \ell^2}{6 P^2}\int_{0}^{\infty} dk \,  \left[ \frac{\log\left( e^{2\pi k} - 1 \right)}{e^{2\pi k}-1} - \frac{\log\left( 1 - e^{-2\pi k} \right)}{1 - e^{-2\pi k}} \right]=-\frac{\pi^2 \ell^2}{36 P^2}.
\end{equation}
Substituting into (\ref{EEtorus}) and using again (\ref{weiersCylinder}) we arrive at
\begin{equation}
    S = \frac{1}{6} \log \left( \frac{P}{\pi \epsilon} \sin \frac{\pi\ell}{P}\right)\,,
\end{equation}
which correctly reproduces the known results for the entanglement entropy of an interval of length $\ell$ on the circle at zero temperature ($P=L$) and on the line at arbitrary temperature ($P=i\beta$)
for any CFT \cite{Korepin:2004zz,Calabrese:2004eu}. Of course, when we also take $P \rightarrow \infty$ we recover the plane result (\ref{entroPlano}).

\subsection{Rényi entropies}\label{sect:5-2}

Proceeding in the same way as in Section \ref{sect:2} it is possible to obtain expressions for more general functions of the reduced density matrix in terms of the resolvent. For instance, the Rényi entropies are given by \cite{Arias:2018tmw}
\begin{equation}
    S_n = \Tr \left[\Theta(G_V-1/2) g_n(G_V)\right] \,,
\end{equation}
with
\begin{equation}
    g_n(\xi) = \frac{1}{n-1} \log\left[ \xi^n - (\xi-1)^n \right]\,. \label{gn}
\end{equation}
Hence we can write
\begin{equation}
    S_n = -\frac{1}{2\pi i} \Tr \ointctrclockwise d\xi \, R(\xi) g_n(\xi). \label{renyitoroF}
\end{equation}
Using the result we obtained for the resolvent, equation (\ref{resolventTorus}), and proceeding in a similar way as for the entanglement entropy, we arrive at the following expression for the Rényi entropies of a chiral scalar on the torus,
\begin{equation}
    S_n = \frac{1+n}{12 n} \log\frac{\sigma\left( \ell \right)}{\epsilon}  + \Delta S_n \label{RenyiFinal}
\end{equation}
with
\begin{equation}\label{DeltaSn}
    \Delta S_n=\frac{1}{2\pi(n-1)} \int_{a}^{b} dx \, \int_{0}^{\infty} dk \, \left[ \nu(x,k) - \nu(x,k+i) \right] \log \frac{1-e^{-2\pi k n}}{(1-e^{-2\pi k})^n}\,.
\end{equation}
This is another example that shows how straightforward the computation of functions of the reduced density matrix can be when we know the resolvent. We have checked that $\Delta S_n$ is real and finite, and that $S_n$ decreases monotonically with $n$, as it should. Of course, in the limit $n\to 1$ equation (\ref{RenyiFinal}) reduces to (\ref{EEtorus}).

As previously mentioned, the problem of determining the resolvent for more than one interval is more challenging. If we could tackle that problem, we would also be able to compute the Rényi mutual information (RMI), a finite quantity that is constructed using the Rényi entropies of two spatial regions $A$ and $B$ by
\begin{equation}
    I_n(A:B)=S_n(A)+S_n(B)-S_n(A\cup B),
\end{equation}
As it is well-known, the entanglement entropy is subadditive and strongly subadditive, which means that the mutual information is positive and monotonic. These properties are in general not shared by the RMI for $n\ne 1$, and we have recently provided 
an
example of non-positive and non-monotonic Rényi entropies in the context of QFT \cite{Blanco:2021kzm}, for the case of the massless Dirac field. It would be nice to see how the RMI behaves in the case of the chiral scalar.

\section{Discussion}\label{sect:6}

In this paper we have computed the entanglement and Rényi entropies of an interval for the chiral scalar on a circle at arbitrary temperature. The key object to obtain these results is the resolvent of the operator (\ref{GV}), which is closely related to the two-point function restricted to pairs of points in the interval. We computed that resolvent by mapping the problem to a sort of Riemann-Hilbert problem, where one looks for an analytic function on the complex torus satisfying a certain jump condition at the interval. This was a challenging problem, but we managed to solve it with ideas that may be useful also for other situations.

The resolvent computed in this paper can also be used to obtain the modular Hamiltonian. Indeed, the modular Hamiltonian of the chiral scalar has the form
\begin{equation}
    H = \int_V dx \int_V dy\, j(x) K(x,y) j(y)\,,\label{Ham}
\end{equation}
and the kernel $K$ is related to the resolvent by
\begin{equation} 
    K=-\frac{i}{2}(\delta')^{-1}\int_{0}^1 d\xi\, R\,. \label{kernelmodham}
\end{equation}
However, this integral is complicated because it has singular terms, namely distributional terms such as delta functions in $x$ and $y$, which make it difficult to compute numerically. Ideally, one would like to extract the singular terms analytically, and then compute numerically the smooth part of $K$. But to do that we need to understand the behavior of the resolvent as $\xi$ approaches the integration limits, and this is not easy in this case because our resolvent involves a complicated function of $\xi$, the function $\mu$ defined in equation (\ref{mu}). Understanding the asymptotic behavior of $\mu$ turns out to be challenging. We are now working on this problem, and we have preliminary results on the asymptotic behavior of $\mu$ that may enable us to compute the modular Hamiltonian in the near future.

We expect the modular Hamiltonian to have a very rich structure. This expectation is based on the results of \cite{Arias:2018tmw} on the modular Hamiltonian of two intervals for the chiral scalar on the plane, which turns out to be remarkably complicated and highly non-local. The intuition coming from the method of images (an intuition made precise in \cite{Blanco:2019cet,Blanco:2019xwi}) suggests that one interval on the torus should be equivalent to an infinite, periodic arrangement of intervals on the plane, so we expect the complicated structure found already for two intervals on the plane to appear even for a single interval on the torus.

The local terms in the modular Hamiltonian are also interesting.
We know that reduced states display local thermal-like properties at high energies, as was shown in \cite{Arias:2016nip}. These thermal properties are characterized by the so-called local temperatures, which are just a generalization of the Unruh temperature associated with
the Rindler wedge. For the vacuum state of two-dimensional models it was argued that the local temperatures are universal \cite{Arias:2016nip}. If this universality were to hold for thermal states as well, then the local temperature in (\ref{kernelmodham}) should be the same as the one for the fermion on the torus, which is given by equation (4.12) in \cite{Blanco:2019cet}.

Another direction in which our work might be extended is by considering more general subregions of the circle, namely unions of intervals. This would be very interesting because, for these more complicated regions, the model exhibits a failure of Haag duality, which leaves an imprint in entanglement quantities such as the mutual information \cite{Arias:2018tmw}.

\acknowledgments

The authors thank Horacio Casini and Markus Fröb for useful discussions. This work has been supported by UBA and CONICET and through the grants PICT 2021-00644, PIP 112202101 00685CO and UBACYT 20020220400140BA. DB also acknowledges support from CONICET through the grant PIBAA 2872021010 0958CO. NA and MK
are supported by CONICET PhD fellowships.

\bibliography{references}
\bibliographystyle{JHEP}

\end{document}